\documentclass[12pt]{article}
\usepackage{amsmath,amssymb,amscd,cite}
\newtheorem{thm}{Theorem}[section]
\newtheorem{prop}[thm]{Proposition}
\newtheorem{lem}[thm]{Lemma}
\newtheorem{cor}[thm]{Corollary}

\newcommand{\pf}{{\bf Proof. \ }}
\newcommand{\qed}{\hfill $\Box$ \\}

\font\msbm=msbm10 at 12pt
\newcommand{\Z}{\mbox{\msbm Z}}

\newcommand{\F}{\mbox{\msbm F}}

\newtheorem{ex}[thm]{Example}

\newcommand{\ord}{ord}

\date{}
\begin{document}
\title{Some Constacyclic Codes over\\Finite Chain Rings}
\author{Aicha Batoul, Kenza Guenda and T. Aaron Gulliver
\thanks{A. Batoul and K. Guenda are with the Faculty of Mathematics USTHB, University
of Science and Technology of Algiers, Algeria.
T. A. Gulliver is with the Department of Electrical and Computer
Engineering, University of Victoria, PO Box 3055, STN CSC, Victoria,
BC, Canada V8W 3P6. email: agullive@ece.uvic.ca.}}
\maketitle

\begin{abstract}

For $\lambda$ an $n$-th power of a unit in a finite chain ring we
prove that $\lambda$-constacyclic repeated-root codes over some
finite chain rings are equivalent to cyclic codes.
This allows us to simplify the structure of some constacylic codes.
We also study the $\alpha +p \beta$-constacyclic codes of length $p^s$
over the Galois ring $GR(p^e,r)$.
\end{abstract}

\section{Introduction}
Constacyclic codes are a generalization of cyclic codes.
While the class of simple root constacyclic codes is well known over
fields, little is known about repeated root constacyclic codes over
fields, and even less about these codes over finite chain rings.
Recently, simple root constacyclic
codes over finite chain rings and the repeated root cyclic and
negacyclic codes over finite fields have been studied \cite{G-G,G-G12}.
In this work, we generalize the results in~\cite{G-G,G-G12}.
The structure of constacyclic
codes over some chain rings is given, as well as conditions on the
equivalence between constacyclic codes over finite chain rings and cyclic codes.
As a special case, we consider the structure of
$(\alpha + \beta p)$ constacyclic codes of length $p^s$ over $GR(p^e,r)$.
Besides contributing to what is known concerning this class of
codes, our motivation in studying these codes comes from the
fact that repeated root cyclic codes contains many optimal
codes, and the decoding complexity can be low as shown by Van Lint
in the binary case~\cite{lint} and Byrne et al.~\cite{byrne} for codes over $Z_4$.
Furthermore, repeated root cyclic codes have found
applications in DNA computing~\cite{watada}.

The remainder of this paper is organized as follows. In Section 2, we
give some preliminaries results concerning finite chain rings.
Section 3 gives conditions on the scalar equivalency between constacylic
codes and cyclic codes over finite fields (for both simple
and repeated roots). In Section 4, we generalize some results of
Section 3 to finite chain rings.
Finally, the structure of $(\alpha + \beta p)$-constacyclic codes of length $p^s$
over $GR(p^e,r)$ is considered in Section 5.

\section{Preliminaries}

A finite chain ring is a finite local, principal commutative ring
$R$ with $1\neq 0$ such that its ideals are ordered by inclusion.
Hence if
$\langle \gamma \rangle$ is the
maximal ideal of the finite chain ring $R$, then $ \gamma $ is
nilpotent with nilpotency index some integer $e$.
The ideals of $R$ form the following chain
$$
<0>= \langle\gamma^e\rangle \subsetneq \langle\gamma^{e-1}\rangle
\subsetneq \ldots \subsetneq \langle\gamma\rangle \subsetneq R.
$$
The simplest examples of finite chain rings are the finite fields.
Since they are chain rings with nilpotency index 0, the nilradical
of $R$ is $\langle\gamma\rangle$, so all the elements of
$\langle\gamma\rangle$ are nilpotent. Therefore the elements of
$R\setminus \langle\gamma\rangle$ are units. We denote this group by
$R^*$. Since $\langle\gamma\rangle$ is a maximal ideal, the residue
ring $\frac{R}{\langle\gamma\rangle}$ is a field which we denote by
$K$. The natural surjective ring morphism is given by
\begin{equation}
\label{eq:over}
\begin{split}
\mu:  R&  \longrightarrow K\\
a &\longmapsto \mu (a)=a +\langle \gamma \rangle
\end{split}
\end{equation}
The map given in (\ref{eq:over}) extends naturally to a map from
$R[x] \longrightarrow K[x]$. A polynomial $f$ of $R[x]$ is called
basic irreducible if $\mu(f)$ is irreducible in $K[x]$.

Let $|R|$ denote the cardinality of $R$. If $|K|=q=p^r$ for some
integer $r$, then
\begin{equation}
\label{cardinality-of-R} |R|=|K|\cdot|\langle\gamma\rangle|=|K|\cdot
|K|^{e-1}=|K|^e=p^{er}.
\end{equation}
A code $C$ of length $n$ over $R$ is a subset of $R$. If the code is
a submodule we say that the code is linear. Here, all codes are
assumed to be linear.

\begin{lem}\label{lem:cardinal2}(\cite{Ana})
Let $R$ be a finite commutative chain ring with maximal ideal
$\langle \gamma \rangle $, residue field $K$ and nilpotency $e$.
Then the following hold:
\begin{enumerate}
  \item [i)] the distinct proper ideals of $R$ are $\langle \gamma^i\rangle$ , $i=1,2,\ldots ,e-1$;
  \item [ii)] for $i=1,2,\ldots ,e-1$ , $|\langle \gamma^i\rangle|\, =\,|K|^{e-i} $.
\end{enumerate}
\end{lem}


A special case of finite chain rings are the called Galois rings.
The Galois ring $GR(p^e,r)$ is a ring of characteristic $p^{e}$ and
cardinality $ p^{er}$. We have $ GR(p^e,1)=\Z_{p^e}$ and $ GR(p,r)=
F_{p^{r}}$. The Galois ring $GR(p^e,r)$ is a local ring with maximal
ideal $\langle p \rangle = p GR(p^e,r)$  and residue field
$GR(p^e,r) /p GR(p^e,r) = F_{p^{r}}$. If $ f(x)\in \Z_{p^e}[x]$ is a
monic basic irreducible polynomial of degree $r$, then the Galois
ring of degree $r$ over $\Z_{p^e}$ is the residue class ring
\[
GR(p^e,r)\,=\,\Z_{p^e}[x] /(f(x)).
\]
If $\xi$ is a root of $f(x)$, then $GR(p^e,r) = \Z_{p^e}[\xi]$,
i.e., $GR(p^e,r)$ is a free module of rank $r$ over $\Z_{p^e}$ with
$\{1,\xi;\xi^{2},\dots,\xi^{r-1} \} $ as a basis. All Galois rings
of the same orders are isomorphic.

There exists an element $\xi$ of order $p^{r}-1 $ in $GR(p^e,r)$ called a primitive
element of  $GR(p^e,r)$,  $\xi$ is a root of a unique monic basic primitive polynomial
of degree $r$ over $\Z_{p^e}$ and dividing $ x^{p^{r}-1}-1$ in $\Z_{p^e}[x]$.
If
$$ \mathfrak{T}_r=\{0,1,\xi,\ldots,\xi^{p^{r}-2}\},$$
then each element $a\,\in GR(p^e,r)$ can be uniquely expressed as
$$a=a_{0}+a_{1}p+\cdots +a_{e-1}p^{e-1}$$
where $ a_{0},a_{1},\ldots,a_{e-1} \in \mathfrak{T}$.
This representation is called the $p$-adic representation of the elements of the Galois ring $GR(p^e,r)$,
and is the generalization of the usual representation of the non-zero elements of a finite field as the powers of a primitive
element.


Let $R$ be a finite chain ring. For a given unit
$\lambda \in R$, a code $C$ is said to be constacyclic, or more
precisely $\lambda-$constacyclic, if $(\lambda c_{n-1}, c_0, c_1,
\ldots, c_{n-2})\in C$, whenever $(c_0, c_1,\ldots, c_{n-1}) \in C$.
The cyclic and negacyclic codes are obtained when $\lambda$ is 1 and
-1, respectively. It is well-known that the $\lambda-$constacyclic
codes over a finite chain ring $R$ are ideals of $R[x]/\langle
x^n-\lambda \rangle$.

Two codes are called equivalent if there exists a monomial
permutation which send one to another.
MacWilliams~\cite{macwilliams} proved that there exists a monomial permutation between two codes over a finite field
if and only if there exists a linear Hamming isometry.
Wood~\cite{wood} extended this result to codes over finite chain rings.
Several weights over rings can be defined.
A weight on a code $C$ over a finite chain ring is called homogeneous
if it satisfies the following assertions.
\begin{itemize}
\item[(i)] $\forall x\in C$ and $\forall u\in R^* : w(x)=w(ux)$
\item[(ii)] There exists a constant $\xi= \xi(w) \in \mathbb{R}$ such that
\[
\sum w(x)_{x\in U} =\xi|U|,
\]
where $U$ is any subcode of $C$.
\end{itemize}
Honold and Nechaev~\cite{nechaev} proved that for codes over a
finite chain ring there exists a homogeneous weight.
A linear morphism $f: R \longmapsto R$ is called a homogeneous isometry if it
is a linear homomorphism which preserve the homogeneous weight.

\begin{lem}\label{lem:greferath}(\cite{greferath})
Let $R$ be a finite chain ring, $C$ a linear code over $R$ and $\phi :
C \longmapsto R^n$ an embedding.
Then the following are equivalent.
\begin{itemize}
\item[(i)] $\phi$ is a homogeneous isometry.
\item[(ii)] $C$ and $\phi(C)$ are equivalent.
\end{itemize}
\end{lem}
Here when two codes are said to be equivalent it means that they are monomially equivalent.

Let $n$ be a positive integer and $q$ a prime power. Then we
denote by $\ord_n(q)$ the multiplicative order of $q$ modulo $n$.
This is the smallest integer $l$ such that $q^l\equiv 1 \bmod n$.
The notation $q\equiv \Box \bmod q$ means that $q$ is a
quadratic residue modulo $n$.

\section{Constacyclic Codes over Finite Fields}

In this section, we give the structure of repeated root constacyclic
codes over finite fields and we give conditions on the existence of
an isomorphism between constacyclic and cyclic codes. We begin with
the following lemma.
\begin{lem}
\label{prop:number} Let $\alpha$ be a primitive element of $\F_{q}$,
$q=p^{r}$ and $\lambda =\alpha^{i}$ for $i \le q-1$. Then the
following holds:
\begin{itemize}
\item[(i)] $x^{n}= \lambda$ has a solution in $\F_{q}$ if and only if $(n,q-1)|i$;

\item[(ii)] if $n=2m$ is an oddly even integer and $q $ is an odd prime power, then
$x^n =-1$ has a solution in $\F_{q}^{*}$ if and only if $-1 \equiv
\Box \bmod q$;

\item[(iii)] $-1 \equiv \Box \bmod q$ if and only if $p\equiv 1 \bmod 4$, $r$ any integer, or $p\equiv 3 \bmod 4$ and $r$ even.
\end{itemize}
\end{lem}
\pf For the Part (i), assume that $x^{n}=\lambda$ has a solution in $\F_q$.
Then this solution is equal to $\gamma=\alpha^j$ for some
$j$ and satisfies $(\alpha^{j})^n=\alpha^{i}$.
This is equivalent to $\alpha^{nj-i}=1$.
Since the order of $\alpha$ is $q-1$, then $(q-1)| nj-i \Leftrightarrow nj \,-\, r(q-1) = i$ for some integer $r$.
This gives that $(n,q-1)|i$.

Assuming the existence of a solution $\alpha^i$ of $x^n=-1$, then
from Part (i) we have that $(n,q-1) | i$. If $n$ is even and $q$ is
odd then $(n,q-1)$ is even, hence $i$ is even. This gives that $-1
=\Box \bmod q$. Conversely, assume that $-1 \equiv \Box \bmod q$.
Then there exists an even $i=2i'$ such that $-1=\alpha^i$. Since
$n=2m$ is oddly even, $(-1)^m=-1=\alpha^{2mi'}=(\alpha^{i'})^n$, and
hence there exists a solution of $x^n+1=0$ in $\F_{q^{r}}$.

For Part (iii), we have that $-1 \equiv \Box \bmod q$ if and only if
$(-1)^\frac{p^r-1}{2}=1 \bmod q$~\cite[Lemma 6.2.4]{huffman03}. This
is equivalent to $p^r \equiv 1 \bmod 4$, which can happen if and
only if $q \equiv 1 \bmod 4$  and $p\equiv 1 \bmod 4$, $r$ any
integer, or $p\equiv 3 \bmod 4$ and $r$ even. \qed
\begin{prop}
\label{lem :2.5} Let $q$ be a prime power, $n$ a positive integer
and $\lambda \in F_q$. If $\F_{q}^{*}$ contains an $n$-th root
$\delta$ of $\lambda$, then a $\lambda$-constacyclic code of length
$n$ is equivalent to a cyclic code of length $n$.
\end{prop}
\pf Let $\delta \in \F_{q}^{*}$ such that $ \delta^{n}\,=\, \lambda$
and define
\[
\begin{tabular}{cccc}

               $ \phi $: $ \F_{q}[x]/(x^{n}-1)$ &$ \longrightarrow$ & $\F_{q}[x]/(x^{n}-\lambda)$ \\
                     $f(x) $& $\longmapsto $& $\phi(f(x))=f(\delta^{-1}x)$ \\
                \end{tabular}
\]
It is obvious that $\phi$ is a ring homomorphism and is Hamming weight preserving.
Hence we only need prove that $\phi$ is a one-to-one map.
For this, let $f(x)$ and $g(x)$ be polynomials in $\F_{q}[x]$
such that $f(x)\equiv g(x) \bmod {x^n-1}.$ This is equivalent to the
existence of $h(x)\in \F_{q}[x]$ such that $f(x)-g(x)=h(x)(x^n-1)$.
This equality is true if and only if $f(\delta^{-1}x)-g(\delta^{-1}x)=
h(\delta^{-1}x)[(\delta^{-1}x)^{n}-1]$ is true.
We have that
$h(\delta^{-1}x)[\delta^{-n}x^{n}-1]=\delta^{-n}h(\delta^{-1}x)[x^{n}-
\delta^{n}]=\delta^{-n}h(\delta^{-1}x)[x^{n}-\lambda]$, so for
$f,g \in \F_{q}[x]/(x^{n}-1)$
\[
\phi(f(x))\, =\, \phi (g(x)) \iff g(x)= f(x).
\]
Then $\phi$ is well defined and one-to-one, and hence is a ring isomorphism and a Hamming isometry.
Since the $\lambda$-constacyclic and cyclic codes are ideals, the result follows from~\cite{macwilliams}.
\qed

\begin{cor}
\label{cor:phi} Let $q=p^r$ be a prime power, $n$ a positive integer and $\lambda \in \F_q$.
If $\F_{q}^{*}$ contains an
$n$-th root $\delta$ of $\lambda$ and the $f_i$, $1 \le i \le r$, are
the monic irreducible factors of $x^{n}-1 $ in $\F_{q}$,
then $x^{n}-\lambda=\lambda \prod_{i=1}^{r}f_{i}(\delta^{-1}x)$ is the
factorization of $x^{n}-\lambda$ into irreducible factors over $\F_{q}$.
\end{cor}
\pf Assume that $x^{n}-1 =\prod_{i=1}^{r}f_{i}(x)$ is the
factorization of $x^{n}-1$ into irreducible factors over $\F_{q}$.
This factorization is unique since it is over a unique factorization
domain (UFD), but in this case $(n,p) \neq 1$, so the factors are not necessarily coprime.
Since $\phi$ is a ring isomorphism,
$\phi(x^n-1)= \delta^{-n}x^{n}-1=\delta^{-n}(x^{n}-\lambda) =
\prod_{i=1}^{r} \phi (f_{i}(x))= \prod_{i=1}^{r}
f_{i}(\delta^{-1}x)$, as the image of $0$ by $\phi$ is $0$. \qed

\begin{thm}
Let $F_q$ be a finite field, $q=p^r$ a prime power, and $n$ an
integer such that $(n,q-1) =1$. Then all constacyclic codes of
length $n$ over $F_q$  are equivalent to cyclic codes of length $n$
over $F_q$.
\end{thm}
\pf Let $\alpha$ be a primitive element of $\F_q$ and $\lambda \in
\F_{q}^*$. Then there exists an integer $i$ such that $\lambda =
\alpha ^{i}$. Since $(n,q-1) =1$, then $(n,q-1)|i$ and by
Lemma~\ref{prop:number} there exists $\delta \in \F_{q}^*$ such that
$\delta ^{n} = \lambda$. By Proposition~\ref{lem :2.5}, the
$\lambda$-constacyclic codes of length $n$ over $\F_{q}$ are
equivalent to cyclic codes of length $n$ over $\F_{q}$. \qed
\begin{ex}
Let $\F_q$ be a finite field, $q=p^r$ a prime power, and $n=mp^{s}$
an integer such that $(m,p^{r}-1)=1$. Then all constacyclic codes of
length $n$ over $\F_q$ are equivalent to cyclic codes of length $n$
over $\F_q$.
\end{ex}
%

\begin{thm}
\label{co:equiv4} Let $\F_q$ be a finite field, $q=p^r$ an odd prime
power, and $m$ an odd integer such that $(m,p)=1$. Let $\lambda \in
\F_{q}^*$ such that there exists $\delta \in \F_{q}^*$ and
$\delta^{m}=\lambda$. Then the following hold:
\begin{enumerate}
 \item [(i)]
  $\pm \lambda$-constacyclic codes of length  $m p^{s}$ over $\F_q$ are  equivalent to cyclic codes over $\F_q$;
  \item [(ii)]
  If $q \equiv 1 \bmod 4$ and $\delta =\beta^{2}$ in $\F_{q}$, then
$\pm \lambda$-constacyclic codes of length $2m p^{s}$ over $\F_q$
are equivalent to cyclic codes over $\F_q$.
\end{enumerate}
\end{thm}

\pf
\begin{enumerate}
  \item [i)]
  Let $\lambda \in \F_{q}^*$ such that there exists $\delta \in \F_{q}^*$ and $\delta^{m}=\lambda$.
  Then exists $\alpha \in \F_{q}^*$ such that $\alpha ^{p^{s}}$ = $\delta$,
  so that $\lambda $ = $ \delta^{m}$ = $\alpha ^{mp^{s}}$.
  Since $mp^{s}$ is odd, we obtain $-\lambda $ = $ (-\delta)^{m}$ = $(-\alpha)^{mp^{s}}$, and the result follows
by Proposition~\ref{lem :2.5}.
  \item [ii)]
  Let $\lambda \in \F_{q}^*$ such that there exists $\beta \in \F_{q}^*$ and $\beta^{2m}=\lambda$.
  Then there exists $\rho \in \F_{q}^*$ so that $\rho ^{p^{s}}$ = $\beta$
  and $\lambda = \beta^{2m}= \rho ^{2mp^{s}}$. Thus $\rho$ is a $2mp^{s}$-th root of $\lambda$ in
  $\F_{q}$.
   If $q \equiv 1 \bmod 4$,
   by Lemma~\ref{prop:number} there exists $\xi \in \F_{q}$ such that $\xi^{2}=-1$.
   Then $-\lambda $ = $ (-1)^{mp^{s}} \beta^{2m}$ = $\xi^{2mp^{s}}\rho^{2mp^{s}}$ = $(\xi \rho) ^{2mp^{s}}$,
   and $\xi\rho$ is a $2mp^{s}$-th root of $-\lambda$ in $\F_{q}$.
   The result then follows by Proposition~\ref{lem :2.5}.
\end{enumerate}
\qed
%
\begin{ex}
When $m$ is odd we have $(-1)^m=-1$. Then from
Part (i) of Theorem~\ref{co:equiv4}, the negacyclic codes of length
$mp^r$ are equivalent to cyclic codes.

When $m$ is odd, $q \equiv 1 \bmod 4$, and $q=p^r$ an odd prime
power, from Part (ii) of Theorem~\ref{co:equiv4} the negacyclic
codes of length $2mp^{s}$ are equivalent to cyclic codes of length
$2mp^{s}$ over $F_{q}$. The structure of these codes was also given
in~\cite{G-G12}, and they are generated by
\[
\prod_{i=1}^{k_1} f_{i}(\beta x)^{\alpha_i} \prod_{i=1}^{k_2}
f_{i}(-\beta x)^{\gamma_i},
\]
where $k_1,k_2 \le l, \alpha_i, \gamma_i \le p^{s}$,
$\prod_{i=1}^{l} f_{i}(x)$ is the factorization of $x^{m}-1$ into
monic irreducible factors over $\F_{q}[x]$, and $\beta^{2}=-1$.
When $m=1$, these codes are generated by $\langle
(x+\beta)^i(x-\beta)^j\rangle$ $ 0\leq i,j \leq p^{s}$. These codes
have also been studied by~Dinh~\cite{permounth3}.
\end{ex}
\begin{ex}
Let $\alpha$ be a primitive element of $\F_{27}^{*}$ and
$n=90=2\cdot5\cdot3^{2}$. Further, let $\lambda \in \{ \alpha ^{2i},
1\leq i \leq 13\}$, and since $(5,26)=1$, $\alpha^{2i} =
\alpha^{2i(26-5\cdot5)} = (\alpha^{-5i})^{2\cdot5}$. Then $\beta
=\alpha^{-5i}$, $\beta^{27}$ = $(\beta^{3})^{9}$ and $\rho =
\beta^{3}$, so $\lambda = (\alpha^{-15})^{2\cdot5\cdot9i}$ =
$(\alpha^{11})^{2\cdot5\cdot9i}$. We have $x^{90}-1 =(x^{10}-1)^9
=(x^5 -1)^9 (x^5 +1)^9= (x-1)^9 (x^4+x^3+x^2+x+1)^9 (x+1)^9
(x^4-x^3+x^2-x+1)^9$. Since cyclic codes of length $90$ over
$\F_{27}$ are principal ideals of $\frac{\F_{27}[x]}{x^{90}-1}$,
these codes are generated by polynomial of the following form
\[
\begin{array}{l}
(x-1)^s (x^4+x^3+x^2+x+1)^j(x+1)^k(x^4-x^3+x^2-x+1)^l\\
\hspace*{0.3in}=
f_{1}^{s}(x)f_{2}^{j}(x)f_{1}^{k}(-x)f_{2}^{l}(-x),$ for $ s,j,k,l
\in \{0,\ldots, 9 \}.
\end{array}
\]
Therefore $\lambda$-constacyclic  codes of length $90$ over
$\F_{27}$ are ideals of $\frac{\F_{27}[x]}{x^{90}-\lambda}$ which is
a principal ideal ring, and these codes are generated by
\[
\begin{array}{l}
\langle f_{1}^{s}(\alpha ^{-11i}x)f_{2}^{j}(\alpha
^{-11i}x)f_{1}^{k}(-\alpha ^{-11i}x)f_{2}^{l}(-\alpha
^{-11i}x)\rangle, \; s,j,k,l \in \{0,\ldots, 9\}, 1 \leq i \leq 13.
\end{array}
\]
\end{ex}
The set of all possible generators of $\lambda$-Constacyclic codes of length 90 over $\F_{27}$ are given in Table 1.

\begin{table}[hbt]
\caption{All Possible Generators of $\lambda$-Constacyclic Codes of length 90 over $\F_{27}$}
\begin{center}
\begin{tabular}{|c|c|}
  \hline
 $\lambda =\alpha^{11.i.2.5.9}$ &$f_{1}^{s}(\alpha ^{-11i}x)f_{2}^{j}(\alpha ^{-11i}x)f_{1}^{k}(-\alpha
^{-11i}x)f_{2}^{l}(-\alpha ^{-11i}x)$
 \\
  \hline
   $\alpha^{11.2.5.9}$  & $\langle f_{1}^{s}(\alpha ^{-11}x)f_{2}^{j}(\alpha ^{-11}x)f_{1}^{k}(-\alpha^{-11}x)f_{2}^{l}(-\alpha ^{-11}x) \rangle$   \\
  $ \alpha^{11.2.2.5.9}$  &$f_{1}^{s}(\alpha ^{-11.2}x)f_{2}^{j}(\alpha ^{-11.2}x)f_{1}^{k}(-\alpha ^{-11.2}x)f_{2}^{l}(-\alpha ^{-11.2}x)$ \\
   $\alpha^{11.3.2.5.9}$& $ f_{1}^{s}(\alpha ^{-11.3}x)f_{2}^{j}(\alpha ^{-11.3}x)f_{1}^{k}(-\alpha
^{-11.3}x)f_{2}^{l}(-\alpha ^{-11.3}x) $  \\
  $ \alpha^{11.4.2.5.9}$ & $f_{1}^{s}(\alpha ^{-11.4}x)f_{2}^{j}(\alpha ^{-11.4}x)f_{1}^{k}(-\alpha
^{-11.4}x)f_{2}^{l}(-\alpha ^{-11.4}x)$ \\
 $  \alpha^{11.5.2.5.9}$& $f_{1}^{s}(\alpha ^{-11.5}x)f_{2}^{j}(\alpha ^{-11.5}x)f_{1}^{k}(-\alpha
^{-11.5}x)f_{2}^{l}(-\alpha ^{-11.5}x)$ \\
$ \alpha^{11.6.2.5.9}$ &$f_{1}^{s}(\alpha ^{-11.6}x)f_{2}^{j}(\alpha
^{-11.6}x)f_{1}^{k}(-\alpha
^{-11.6}x)f_{2}^{l}(-\alpha ^{-11.6}x)$ \\
  $ \alpha^{11.7.2.5.9}$ & $ f_{1}^{s}(\alpha ^{-11.7}x)f_{2}^{j}(\alpha ^{-11.7}x)f_{1}^{k}(-\alpha
^{-11.7}x)f_{2}^{l}(-\alpha ^{-11.7}x)$\\
 $ \alpha^{11.8.2.5.9}$  & $f_{1}^{s}(\alpha ^{-11.8}x)f_{2}^{j}(\alpha ^{-11.8}x)f_{1}^{k}(-\alpha
^{-11.8}x)f_{2}^{l}(-\alpha ^{-11.8}x)$ \\
 $ \alpha^{11.9.2.5.9}$ &$f_{1}^{s}(\alpha ^{-11.9}x)f_{2}^{j}(\alpha ^{-11.9}x)f_{1}^{k}(-\alpha
^{-11.9}x)f_{2}^{l}(-\alpha ^{-11.9}x)$ \\
$ \alpha^{11.10.2.5.9}$&$f_{1}^{s}(\alpha
^{-11.10}x)f_{2}^{j}(\alpha ^{-11.10}x)f_{1}^{k}(-\alpha
^{-11.10}x)f_{2}^{l}(-\alpha ^{-11.10}x)$ \\
$ \alpha^{11.11.2.5.9}$ & $f_{1}^{s}(\alpha
^{-11.11}x)f_{2}^{j}(\alpha ^{-11.11}x)f_{1}^{k}(-\alpha
^{-11.11}x)f_{2}^{l}(-\alpha ^{-11.11}x)$ \\
$ \alpha^{11.12.2.5.9}$  & $f_{1}^{s}(\alpha
^{-11.12}x)f_{2}^{j}(\alpha ^{-11.12}x)f_{1}^{k}(-\alpha
^{-11.12}x)f_{2}^{l}(-\alpha ^{-11.12}x)$ \\
 $ \alpha^{11.13.2.5.9}$ & $f_{1}^{s}(\alpha ^{-11.13}x)f_{2}^{j}(\alpha ^{-11.13}x)f_{1}^{k}(-\alpha
^{-11.13}x)f_{2}^{l}(-\alpha ^{-11.13}x)$ \\
  \hline
\end{tabular}
\end{center}
\end{table}
\section{Constacyclic Codes over Finite Chain Rings}

The purpose of this section is to provide an isomorphism between
constacyclic and cyclic codes in a more general setting than those
given in \cite{aicha2011}. This allows us to simplify the structure
of constacyclic codes. To achieve this goal, we extend some results
given in the previous section to finite chain rings with
characteristic $p$ such that $(n,p)=1$. For $\lambda \in R^*$, if
there exists $ \delta \in R^*$ such that $\delta^{n}=\lambda$, then
$\delta$ is called an $n$-th root of $\lambda$ in $R$.

Since $R^*=R\setminus \langle \gamma \rangle$, then for all $a \in R^*$ we have $\mu (a)\neq 0$.
Further, as the map $\mu$ is a surjective homomorphism, it
induces the surjective homomorphism $\tilde{\mu} : R^{*} \longrightarrow \F_q^{*}$.
The following lemma provides the link between the $n$-th root of elements in $R$ and the $n$-th root of their images in $\F_q$.
\begin{lem}(\cite[p. 355]{Mac})
\label{lem:dough} With the notation given above, $Ker(\tilde{\mu})$ is a $p$-group.
%
\end{lem}
We now generalize a result of Dougherty et al.~\cite[Lemma
4.2]{D-L}.
\begin{prop}
\label{lem :2.7} Let $R$ be a finite chain ring  with residue field
$\F_{p^{r}}$ and $n$ an integer such that $(p,n)=1$. Further, let
$\lambda \in R$ such that $\mu(\lambda)\neq 0$.
Then there exists an $n$-th root of $\lambda$ in $R$ if and only if there exists an
$n$-th root of $\mu(\lambda)$ in $\F_{q}^{*}$.
\end{prop}
\pf We first prove that the following map is an automorphism $\xi
:Ker(\tilde{\mu}) \longrightarrow Ker(\tilde{\mu}), x \longmapsto
x^{n}.$ Suppose $x,y \in Ker(\tilde{\mu})$ are such that
$x^{n}=y^{n}$, then $(xy^{-1})^{n}=1$. From Lemma~\ref{lem:dough}, we
have that $Ker(\tilde{\mu})$ is a $p$-group, and since $(n,p)=1$ we
obtain that $xy^{-1}=1$. Hence $x=y$, which implies that $\xi$ is a bijection.
It is easy to check that $\xi$ is a homomorphism, and hence an automorphism.
Let $\lambda\in R $ with $\mu(\lambda) \neq 0$, and suppose $\exists \delta \in R$ such that $ \delta^{n}=\lambda$.
Then since $\mu$ is a homomorphism, we have
$\mu(\delta^{n})=\mu(\lambda)\Longrightarrow \mu(\lambda)=(\mu(\delta))^{n}$.
Now let $c_{1}\in \F_{p^{r}}^{*}$ such that there exists $\delta_{1}\in \F_{p^{r}}^{*}$ which satisfies
$c_{1}=\delta_{1}^{n}$. Then there exists $c_{0}\in R^{*}$ and
$\delta_{0}\in R^{*}$ such that $\mu(c_{0})=c_{1}$ and
$\mu(\delta_{0})=\delta_{1}$.
Therefore $\mu(c_{0})=\mu(\delta_{0}^{n})$, so then $c_{0}\delta_{0}^{-n} \in Ker(\tilde{\mu})$.
Since we already have that $\xi$ is a bijection, there exists $b\in Ker(\tilde{\mu})$ such that
$c_{0}\delta_{0}^{-n}=b^n$.
Then $c_{0} = \delta_{0}^{n}b^{n} = (\delta_{0} b)^{n}$, $\delta_{0} b \in R^{*}$.
This implies that $\delta_{0} b$ is a $n$-th root of $c_{0}$ in $R^{*}$.
\qed
\begin{thm}
\label{th:main} Let $R$ be a finite chain ring with residue field
$\F_{q}$, and let $n$ be an integer such that $(n,q)=1$. If there
exists $\lambda \in F_{q}^{*}$ such that $\exists \delta \in
\F_{q}^{*}$ with $\delta^{n}=\lambda$, then $\exists \lambda_0$ a
unit in $ R $ such that $\lambda _0= {\delta _0}^n$ and $\mu(\lambda_0)=\lambda $.
Further, the $\lambda_0$-constacyclic code of length $n$ over $R$ is equivalent to
a cyclic code of length $n$ over $R$.

Conversely, if there exists a unit $\lambda\in R$ such that there
exists a unit $\delta \in R$ with $\delta^{n}=\lambda$, then
a $\mu(\lambda)$-constacyclic code of length $n$ over $\F_{q}$ is
equivalent to a cyclic code of length $n$ over $\F_{q}$.
\end{thm}
\pf Let $R$ be a finite chain ring with residue field $\F_{q}$, and
let $n$ be an integer such that $(n,q)=1$. If there exists $\lambda
\in F_{q}^{*}$ such that $\exists \delta \in \F_{q}^{*}$ with $\delta^{n}=\lambda$,
then by Proposition \ref{lem :2.7} there exists $\lambda_0 ,\delta_0 \in R^*$
such that $\mu(\lambda_0)=\lambda $ and $\delta_0^n = \lambda_0$.
Let
\begin{equation}
\label{eq:isomo}
\begin{tabular}{cccc}
               $ \psi $: $ R[x]/(x^{n}-1)$ &$ \longrightarrow$ & $R[x]/(x^{n}-\lambda_0)$ \\
                    $f(x) $& $\longmapsto $& $\mu(f(x))=f(\delta_0^{-1}x)$. \\
                    \end{tabular}
\end{equation}
We need to prove that $\psi$ is an isometry according to a
homogeneous weight over $R$. Let $w(.)$ be a homogeneous weight over
$R$ and let $f(x)=a_0+a_1x+\ldots a_nx^n$ be a codeword in
$R[x]/(x^{n}-1)$. Then $\psi(f(x))=a_0+a_1\delta^{-1}x+a_2
{\delta^{-1}x}^2 +\ldots+{\delta^{-1}x}^n$.
Since $\delta$ is a unit, it must be that $w(\delta^{-i}a_i)=w(a_i)$, hence $w(\psi(f(x))=w(f(x))$.
By a proof similar to that of Proposition \ref{lem :2.5} for fields,
$\psi$ is a ring isomorphism so $A$ is an ideal of $R[x]/(x^{n}-1)$.
Then $\psi(A)$ is an ideal of $R[x]/(x^{n}-\lambda_0)$ and if $B$ is an ideal of $R[x]/(x^{n}-\lambda_0)$,
$\psi^{^-1}(B)$ is an ideal of $ R[x]/(x^{n}-1)$.
Hence from Lemma~\ref{lem:greferath} the result follows.

Conversely, if there exists a unit $\lambda\in R$ such that there
exists a unit $\delta \in R$ with $\delta^{n}=\lambda$, then by
Proposition~\ref{lem :2.7} $\mu(\lambda) = \mu(\delta)^n$ in $F_q$.
From Proposition~\ref{lem :2.5}, we deduce that a
$\mu(\lambda)$-constacyclic code of length $n$ over $\F_{q}$ is
equivalent to a cyclic code of length $n$ over $\F_{q}$. \qed
\begin{cor}
\label{cor:psi} Let $R$ be a finite chain ring with residue field
$F_q$ with $q=p^r$ a prime power and $n$ a positive integer such that $(n,p)=1$.
In addition, let $\lambda $ be a unit in $R$.
Then if $R$ contains an $n$-th root $\delta$ of $\lambda$ and the
$f_i$, $1 \le i \le r$ are the monic basic irreducible pairwise
coprime factors of $x^{n}-1 $ in $R[x]$, $x^{n}-\lambda=\lambda
\prod_{i=1}^{r}f_{i}(\delta^{-1}x)$ is the factorization of
$x^{n}-\lambda$ into basic irreducible pairwise coprime factors in $R[x]$.
\end{cor}
\pf Assume that $x^{n}-1 =\prod_{i=1}^{r}f_{i}(x)$ is the unique
factorization of $x^{n}-1$ into monic basic irreducible pairwise coprime factors over $R$.
This factorization is unique since $x^{n}-1$ is monic, $(n,p)=1$ and it is over a finite chain ring.
Since $\mu$ is a ring homomorphism, $\mu (x^{n}-1) =\prod_{i=1}^{r}\mu (f_{i})(x)$ is the unique factorization
of $\mu (x^n-1)$ into monic irreducible pairwise coprime factors over $F_q$.
By Corollary~\ref{cor:phi}, $x^n-\mu(\lambda)$ = $\mu(\lambda) \prod_{i=1}^{r}\mu(f_{i})(\mu (\delta)^{-1}x)$ is the unique
factorization of $x^n-\mu(\lambda)$ into irreducible pairwise coprime over $F_q$.

From Theorem~\ref{th:main}, $\psi$ is a ring isomorphism, so $\psi(x^n-1)=
\delta^{-n}x^{n}-1=\delta^{-n}(x^{n}-\lambda) = \prod_{i=1}^{r} \psi
(f_{i}(x))= \prod_{i=1}^{r} f_{i}(\delta^{-1}x)$, and $\lambda
\prod_{i=1}^{r} f_{i}(\delta^{-1}x)$ is a factorization of $x^{n}-\lambda$ over $R$.
Then $\mu(x^{n}-\lambda) = \mu(\prod_{i=1}^{r} f_{i}(\delta^{-1}x)) = \prod_{i=1}^{r} \mu (f_{i})(\delta^{-1}x) =
\prod_{i=1}^{r} \mu (f_{i})(\mu(\delta)^{-1}x)$,
$ f_{i}(\delta^{-1}x)$ is basic irreducible over $R$ for $1\leq i\leq r$
\qed
\begin{cor}
Let $R$ be a chain ring with maximal ideal $\langle\gamma \rangle$,
index of nilpotency $e$, and residue field $\F_{q}$ with $q = p^{r}
$ a prime power. Further, let $n$ be an integer such that $(n,q)=1$
and $\lambda \in 1+ \langle\gamma \rangle$. Then
$\lambda$-constacyclic codes over $R$ of length $n$ are equivalent
to cyclic codes of length $n$ over $R$.
 \end{cor}
 \pf
Let $\lambda \in 1+ \langle \gamma\rangle$ so that $\mu( \lambda)=1$.
Then from Proposition~\ref{lem :2.7}, there exists an $n$-th root of
$\lambda$ in $R$ since there is an $n$-th root of $1$ in $\F_q$.
Hence from Theorem~\ref{th:main}, we have that these constacyclic
codes are equivalent to cyclic codes.
\qed
\begin{ex}
\label{cor:jit} If $n$ is an integer such that $(n,q)=1$ and
$\lambda =1\pm \gamma^{e-1}$, then $\lambda$-constacyclic codes over
$R$ of length $n$ are equivalent to cyclic codes of length $n$ over
$R$.
 \end{ex}
\begin{cor}
\label{th:main4} Let $R$ be a finite chain ring with residue field
$\F_{p^{r}}$, $n=2m$ an oddly even integer and $p$ an odd prime such
that $(n,p)=1$. Then if $p\equiv 1 \bmod 4$, $r$ an integer, or
$p\equiv 3 \bmod 4$, $r$ even, negacyclic codes of length $n$ over
$R$ are equivalent to cyclic codes of length $n$ over $R$.
\end{cor}
\pf From Lemma~\ref{prop:number}, if $n$ is oddly even and $p\equiv
1 \bmod 4$, $r$ an integer, or $p\equiv 3 \bmod 4$, $r$ even, then $x^n+1$ has a solution in $\F_{p^r}$.
Hence from Theorem~\ref{th:main} the negacyclic codes are equivalent to cyclic codes over $R$.
\qed

Next the structure of negacyclic codes of length $2m$ are examined
under the assumptions of Corollary~\ref{th:main4}.
\begin{prop}
With the assumptions of Corollary~\ref{th:main4}, any cyclic or
negacyclic code over $R$ of length $2m$, $m$ odd, over $R$ is a
direct sum of two cyclic codes of length $m$ over $R$.
\end{prop}
\pf Let  $n=2m $ with $m$ an odd integer  and $p\equiv 1 \bmod 4$,
$r$ an integer, or $p\equiv 3 \bmod 4$, $r$ even, by
Lemma~\ref{prop:number} there exists $\nu \in F_q^*$ such that
$\nu^2=-1$. Then since $p$ is odd and $(p,2)=1$, by
Proposition~\ref{lem :2.7} there exists a unit $\nu_0 \in R $ such
that $\nu_0^2 =-1$ in $R$. Thus  $-1$ = $ (-1)^m$ = $(\nu_0^2)^m$ =
$(\nu_0)^{2m}$ = $\nu_0^n$. From Theorem~\ref{th:main} we deduce
that negacyclic codes of length $n$ are equivalent to cyclic codes
of length $n$ over $R$. Since $m$ is odd, then
$\frac{R[x]}{x^{m}-1}\simeq \frac{R[x]}{x^{m}+1}$. Hence we have
\[
\frac{R[x]}{x^{2m}-1} \simeq \frac{R[x]}{x^{m}-1}\oplus
\frac{R[x]}{x^{m}-1}.
\]
\qed

\begin{ex}
Let $R=\Z_{25}$ and $n=18=2\cdot9$. Since $x^{18}-1=(x^9-1)(x^9+1)$, then
\[
\frac{\Z_{25}[x]}{x^{18}-1} \simeq \frac{\Z_{25}[x]}{x^{9}-1}\oplus
\frac{\Z_{25}[x]}{x^{9}+1}.
\]
In $\Z_{25}[x]$ we have the factorization $x^9-1=(x+24)(x^2+x+1)(x^6+x^3+1)$.
Denote the monic basic irreducible factors of $x^9-1$ as
\[
f_{0}(x) = (x+24),\, f_{1}(x)=(x^2+x+1),\text{ and }
f_{2}(x)=(x^6+x^3+1).
\]
Then
\[
x^{18}-1 =(x^9-1) (x^9+1) =
f_{0}(x)f_{1}(x)f_{2}(x)f_{0}(-x)f_{1}(-x)f_{2}(-x)
\]
is the factorization of $x^{18}-1$ in $ \Z_{25}[x]$ into monic basic
irreducible factors.
Then
\[
C_{1}= \langle f_{0}(x)f_{2}(x),5f_{0}(x)f_{1}(x) \rangle
\]
and
\[
C_{2}=\langle f_{0}(x)f_{1}(x),5f_{1}(x)f_{2}(x)\rangle
\]
generate cyclic codes of length $9$ over $\Z_{25}$.
Since $9$ is odd, $\frac{\Z_{25}[x]}{x^{9}-1}\simeq \frac{\Z_{25}[x]}{x^{9}+1}$,
and the image of $C_{2}$ is $ \langle f_{0}(-x)f_{1}(-x),5f_{1}(-x)f_{2}(-x)\rangle$.
Then
\[
\langle f_{0}(x)f_{2}(x),5f_{0}(x)f_{1}(x) \rangle \oplus  \langle
f_{0}(-x)f_{1}(-x),5f_{1}(-x)f_{2}(-x)\rangle
\]
is a cyclic code of length $18$ over $\Z_{25}$.

We know that negacyclic codes over $\Z_{25}$ of length $18$ are
principal ideals of $\frac{\Z_{25}[x]}{x^{18}+1}$.
In $\Z_{25}$, $7^2=-1$ so $x^{18}+1= x^{18}-(-1)=x^{18}-7^2=x^{9}-7)(x^{9}+7) =
(x^{9}-7^9)(x^{9}+7^9)= 7((-7x)^{9}-1)7((7x)^{9}+1) =
-((-7x)^{9}-1)((7x)^{9}+1)$.
Since $x^9-1=f_{0}(x)f_{1}(x)f_{2}(x)$, $f_{0}(-7x)f_{1}(-7x)f_{2}(-7 x)$ is the
factorization of $((-7x)^{9}-1)$ into monic basic irreducible
factors in $\Z_{25}[x]$, and $f_{0}(7x)f_{1}(7x)f_{2}(7 x)$ is the
factorization of $((7x)^{9}+1)$ into monic basic irreducible
factors in $ \Z_{25}[x]$.
We then have
\begin{equation}
 \frac{\Z_{25}[x]}{x^{18}+1}
\simeq \frac{\Z_{25}[x]}{(-7x)^9-1}\oplus
\frac{\Z_{25}[x]}{(7x)^9+1}.
\end{equation}
Since $\frac{\Z_{25}[x]}{(7x)^{9}-1}\simeq
\frac{\Z_{25}[x]}{(7x)^{9}+1}$, then for example
\[
\langle f_{0}(7x)f_{2}(7x),5f_{0}(7x)f_{1}(7x) \rangle \oplus
\langle f_{0}(-7x)f_{1}(-7x),5f_{1}(-7x)f_{2}(-7x)\rangle
\]
is a negacyclic code of length $18$ over $\Z_{25}$.
\end{ex}

\section{$(\alpha+\beta p )$-Constacyclic Codes of Length $p^s$ over Finite Galois Rings}

Let $R= GR(p^e,r)$ be the finite Galois ring with residue field $\F_{p^r}$,
maximal ideal $\langle p \rangle$, and nilpotency index $e$.
Let $\alpha,\beta$ be units in $R$, $n = p^{s}$, and
$\mathcal{R}(\alpha,\beta)= \frac{R[x]}{(x^{p^{s}}-(\alpha +\beta p))}$.
Then the $(\alpha+\beta p)$-constacyclic codes of length $p^s$ over
$R$ are precisely the ideals of $\mathcal{R}(\alpha,\beta)$.

Each element of $GR(p^e,r)$ can be uniquely written as
$$a=a_0+a_1p+a_2p^2+\cdots +a_{e-1}p^{e-1}$$ with $a_i \in \mathcal{T}$.
Note that $\mathcal{T}$ is equivalent to $ F_{p^r}$.
We have $\mu(a)= a_0$, where $\mu$ is the map given by~(\ref{eq:over}).

\begin{lem}\label{nilpotent}
Let $R$ be a finite commutative ring with identity, and let $x,y \in R$.
If $x-y$ is nilpotent in $R$, then $x$ is a unit if and only if $y$ is a unit.
\end{lem}
\pf Let $z=x-y$.
Since the set of nilpotent element of $R$ is a
subgroup in $R$, if $z$ is nilpotent then $x$ is nilpotent if and
only if $y$ is nilpotent.
\qed

\begin{lem}
\label{lem:p unit} Assume there exists a unit $\alpha_{0} \in R $
such that $\alpha_{0}^{p^s} = \alpha $. Then in
$\mathcal{R}(\alpha,\beta)$ we have $(\alpha_0^{-1} x -1)^{p^{s}}= p
\rho(x)$ where $\rho(x)$ is a unit in $\mathcal{R}(\alpha,\beta)$.
Moreover, the nilpotency index of $(\alpha_0^{-1} x -1)^{p^{s}}$ is $ep^s$.
\end{lem}
\pf In $\mathcal{R}(\alpha,\beta)$ we have
\[
\begin{array}{ccl}
(\alpha_{0}^{-1} x -1)^{p^s}&=&(\alpha_0^{-1} x)^{p^{s}}+(-1)^{p^{s}} + \sum_{i=1}^{p^s-1}(-1)^i(^{p^s}_{i}) (\alpha_0^{-1} x)^{p^{s}-i}\\
&=&\alpha_0^{-p^s} x^{p^{s}}+(-1)^{p^{s}} + \sum_{i=1}^{p^s-1}(-1)^i(^{p^s}_{i}) (\alpha_0^{-1} x)^{p^{s}-i}\\
&=&\alpha ^{-1} ( \alpha +\beta \gamma )+(-1)^{p^{s}} + \sum_{i=1}^{p^s-1}(-1)^i(^{p^s}_{i}) (\alpha_0^{-1} x)^{p^{s}-i}\\
&=&1+(-1)^{p^{s}}+\alpha^{-1}\beta p
+\sum_{i=1}^{p^s-1}(-1)^i(^{p^s}_{i}) (\alpha_0^{-1} x)^{p^{s}-i}
\end{array}
\]
Let $g(x)=\sum_{i=1}^{p^s-1}(-1)^i(^{p^s}_{i}) (\alpha_0^{-1}x)^{p^{s}-i}$.
Expanding $g(x)$ in $(\alpha_0^{-1} x-1)$ gives
\[
\begin{array}{ccl}
g(x)&=&\sum_{i=1}^{p^s-1}(-1)^i(^{p^s}_{i}) ((\alpha_0^{-1} x-1)+1)^{p^{s}-i}\\
&=&\sum_{i=1}^{p^s-1}\sum_{j=0}^{p^s-i}(-1)^i(^{p^s}_{i})(^{p^s-i}_{j})
(\alpha_0^{-1} x-1)^{p^{s}-i-j},
\end{array}
\]
where the constant term is
\[
g(\alpha_0) = (-1)-(-1)^{p^{s}}.
\]
Hence $g(x)$ can be represented as $g(\alpha_0) +
p\sum_{i=1}^{p^s-1}b_i(\alpha_0^{-1} x-1)^i$ where $b_i \in R $ for
$0\leq i\leq p^s-1$. Then we have $(\alpha_0^{-1}
x-1)^{p^{s}}=p\rho(x)$, where
\[
\rho(x)=\alpha^{-1}\beta +\sum_{i=1}^{p^s-1}b_i(\alpha_0^{-1}
x-1)^i.
\]
From Lemma~\ref{lem:rho}, we have that $\rho(x)$ is a unit in
$\mathcal{R}(\alpha,\beta)$ since $\alpha^{-1}\beta $ is a unit in $R$.
Hence $(\alpha_0^{-1} x-1)^{p^se}=0$ in $\mathcal{R}(\alpha,\beta)$,
which means that the nilpotency index of $(\alpha_0^{-1} x-1)$ is $ep^s$.
\qed

\begin{lem}
\label{lem:rho} Assume there exists a unit $\alpha_{0} \in R$ such
that $\alpha_{0}^{p^s} = \alpha$. Then for $f(x)\in
\mathcal{R}(\alpha,\beta)$ the following hold:
\begin{enumerate}
\item [(i)] $f(x)$ can be uniquely written as
\[
f(x)= a_0 + a_1( \alpha_{0}^{-1} x -1 )+ a_2( \alpha_{0}^{-1} x -1
)^{2}+\cdots +a_{p^s-1}( \alpha_{0}^{-1} x -1 )^{p^s-1}
\]
where $a_i \in R , 0\leq i\leq p^s-1$;
\item [(ii)] $f(x)$ is a unit in $\mathcal{R}(\alpha,\beta)$ if and only if $\mu(a_0)\neq 0$.
\end{enumerate}
\end{lem}
\pf Note that we can write $x$ as $x= \alpha_0 ( \alpha_{0}^{-1} x-1)+ \alpha_0$
so that $x^{i}= (\alpha_0 ( \alpha_{0}^{-1} x -1 )+\alpha_0)^{i}$ for $0\leq i\leq p^s-1$.
Hence $f(x)\in \mathcal{R}(\alpha,\beta) $ can be written as
\[
f(x)= a_0 +a_1( \alpha_{0}^{-1} x -1 )+ a_2( \alpha_{0}^{-1} x -1 )^{2}+\cdots
+a_{p^s-1}( \alpha_{0}^{-1} x -1 )^{p^s-1}.
\]
Since $a_0 \in R$, it can be written uniquely as $a_0=\mu(a_0) +a_{0,1} p + \cdots +
a_{0,e-1}p^{e-1}$ with $a_{0,i} \in F_{p^r}$.
Hence $f(x)$ can be expressed as
\[
f(x) = \mu(a_0)+ a p  + (\alpha_0^{-1} x -1)g(x),
\]
for some $a \in R $ and $g(x) \in \mathcal{R}(\alpha,\beta)$.
Define
\[
A(x) = f(x) - \mu(a_0).
\]
Since $(\alpha_0^{-1} x -1)$ and $p$ are nilpotent in
$\mathcal{R}(\alpha,\beta)$, it follows that $a p$ is nilpotent in $\mathcal{R}(\alpha,\beta)$.
Therefore $A(x)$ is nilpotent in $\mathcal{R}(\alpha,\beta)$, so by Lemma~\ref{nilpotent} $f(x)$ is
a unit if and only if $\mu(a_0)$ is a unit in $F_{p^r}$, i.e., $\mu(a_0)\neq 0 $.
\qed

\begin{thm}
\label{thm:Mai} The ring $\mathcal{R}(\alpha,\beta)$ is a chain ring
with maximal ideal $\langle(\alpha_0^{-1} x-1)\rangle$ with residue
field $F_{p^r}$, and the nilpotency index of $(\alpha_0^{-1} x-1)$ is $ep^s$.
The ideals of $\mathcal{R}(\alpha,\beta)$ are
$\langle(\alpha_0^{-1} x-1)^{i}\rangle$, $0\leq i\leq e p^s$.
\end{thm}
\pf Let $a(x)$ be an element in $\mathcal{R}(\alpha,\beta)$. Then
from Lemma~\ref{lem:rho}, $a(x)$ can be expressed as
\[
a(x) = a_0+ pa +(\alpha_0^{-1}x-1)g(x),
\]
where $a_0 \in F_{p^r}$, $a \in R $ and $g(x)\in \mathcal{R}(\alpha,\beta)$.
If $a_0=0$, then $a(x) = p a +(\alpha_0^{-1}x-1)g(x)$.
By Lemma~\ref{lem:p unit}, $p =(\alpha_0^{-1} x-1)^{p^{s}} (\rho(x))^{-1}$,
and hence $a(x) = (\alpha_0^{-1}x-1)h(x)$ for some $h(x)$ in $\mathcal{R}(\alpha,\beta)$.
This gives that $a(x) \in \langle(\alpha_0^{-1}x-1)\rangle$.
If $a_0 \neq 0$, then $a(x)$ is a unit in $\mathcal{R}(\alpha,\beta)$.
Therefore for any element $a(x)\in \mathcal{R}(\alpha,\beta)$ either
$a(x)$ is a unit or $a(x)\in\langle(\alpha_0^{-1}x-1)\rangle$.
This implies that $\mathcal{R}(\alpha,\beta)$ is a local ring  with maximum
ideal $\langle(\alpha_0^{-1}x-1)\rangle$.
Hence from \cite[Proposition 2.1]{permounth}, $\mathcal{R}(\alpha,\beta)$ is a chain ring
whose ideals are $\langle(\alpha_0^{-1} x-1)^{i}\rangle$, $0\leq i\leq e p^s$.
\qed

\begin{cor}
There are $1+ep^s$ $(\alpha+\beta p)$-constacyclic codes of length $p^s$ over $R$.
They are precisely the ideals $C_i=\langle(\alpha_0^{-1} x-1)^{i}\rangle
\subset\mathcal{R}(\alpha,\beta)$ for some $0\le i \le ep^s$.
Then the number of codewords in $C_i$  is $|C_i| = p^{r(ep^s-i)}$.
\end{cor}
\pf An $(\alpha+\beta \gamma)$-constacyclic code of length $p^s$ over
$R$ is an ideal of $\mathcal{R}(\alpha,\beta)$.
From Theorem~\ref{thm:Mai}, we have that these ideals are
$C_i=\langle(\alpha_0^{-1} x-1)^{i}\rangle$ with $0\le i \le ep^s$.
Then by Lemma~\ref{lem:cardinal2}, we deduce that $|C_i|\, =\, p^{r(ep^s-i)}$.
\qed
\begin{ex}
Let $\Z_9 =\{0,3,6\}\cup \{2,2^2,2^3,2^4,2^5,2^6\}$ and $n=3^3$. We
have that $(3^3,6)=3$ if $\alpha = 2^3$. Let $\beta$ be a unit in
$Z_9$. If $C$ is a $(-1+3\beta)$-constacyclic code over $Z_9$ of
length $27$, then $C=\langle (-x-1)^i\rangle$ for some $i\in \{0,1,
\ldots , 2 \cdot 3^3\}$, and the number of codewords in $C$ is $|C|=
3^{2\cdot27-i}$.
\end{ex}

\end{document}